\documentstyle[prl,aps,epsf,twocolumn]{revtex}

\tighten

\begin{document}
\twocolumn[\hsize\textwidth\columnwidth\hsize\csname @twocolumnfalse\endcsname

\title{\bf Elementary mechanisms governing the dynamics of silica}

\author{Normand Mousseau$\ast,\dagger$, G. T. Barkema$\ddagger$ and
Simon W. de Leeuw$\dagger$}

\address{ $\ast$  Department of Physics and Astronomy, Ohio University,
Athens, OH 45701, USA\\
(permanent address) }

\address{
$\dagger$ Computational Physics, Dept. of Applied Physics, TU Delft,
Lorentzweg 1, 2628 CJ Delft,\\
the Netherlands}

\address{
$\ddagger$Theoretical Physics, Utrecht University, Princetonplein 5,
3584 CC Utrecht, the Netherlands}

\date{\today}

\maketitle

\begin{abstract}
A full understanding of glasses requires an accurate atomistic picture
of the complex activated processes that constitute the low-temperature
dynamics of these materials. To this end,
we generate over five thousand activated events
in silica glass, using the activation-relaxation technique;  these
atomistic mechanisms are analysed and classified according to their
activation energies, their topological properties and their spatial
extend. We find that these are collective processes, involving ten to
hundreds of atoms with a continuous range of activation energies; that
diffusion and relaxation occurs through the creation, annihilation and
motion of single dangling bonds; and that silicon and oxygen have
essentially the same diffusivity.
\end{abstract}
\vspace*{1.0cm}
]

\narrowtext

\newpage
Glassiness is a dramatic slowing down of the kinetics of a liquid
as the temperature decreases below some typical value.  Experiments
have yielded considerable information about the macroscopic character
of this phenomenon, but very few techniques provide the local probe
needed to understand its  microscopic
origin~\cite{griscom86,limoge90,lamkin92}. On the theoretical side,
significant progress has been made recently in understanding
the supercooled region, but little is known about the atomistic
nature of the relaxation and diffusion dynamics taking place at
temperatures below the glass transition~\cite{stillinger98}.  Using a
new Monte Carlo technique, the activation-relaxation technique, we map
in detail the activated processes of {\it g}-SiO$_2$ taking place at
low temperatures. 

\begin{figure}
\vspace{1cm}

\epsfxsize=7.5cm
\epsfbox{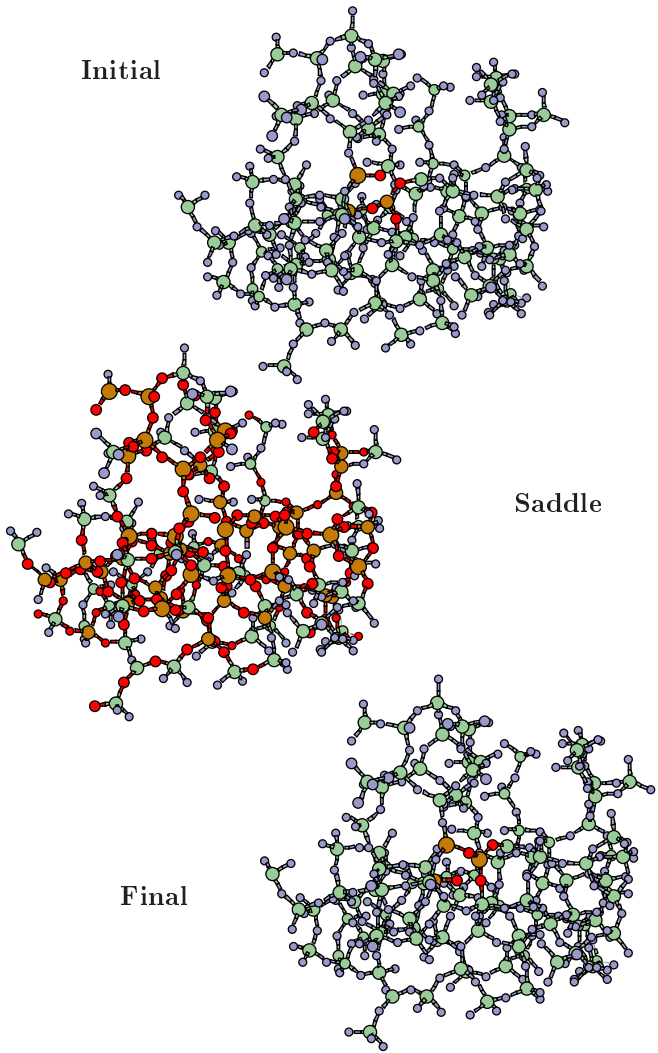}

 \vspace{3cm}
\caption{ 
An event takes place in two stages: the activation from the initial
minimum (top) to a saddle point (middle), and the relaxation from 
there to the final minimum (bottom). For clarity, we have depicted 
only the 187 atoms that move more than 0.1 \AA, plus their nearest 
neighbors.  Large circles are Si atoms, small ones O. 
In the top and bottom figures, the few atoms that are actually
involved in the change of topology, plus their neighbors, are
identified by a different color; this
particular event is the hopping of a dangling bond on an O to a near
neighbor.  In the middle figure, the different coloring separates the
atoms involved from their immobile neighbors.
\label{fig:danglingdiff}}
\end{figure}

The activation-relaxation technique (ART) is a method that allows an
efficient sampling of activated processes (events) in complex
continuous systems~\cite{art,artb}.  Moves are defined directly in the
configurational energy landscape and can reach any level of complexity
required by the dynamics; they can involve hundreds of atoms crossing
barriers as high as 25 eV. In a two-step process, a configuration is
first brought from a local minimum to an adjacent saddle point and then
relaxed to a new minimum. Such an event is shown in figure
\ref{fig:danglingdiff}.  Each event is accepted or rejected following a
standard Metropolis procedure.  In this work, we study two independent
runs on 1200-atom cells of SiO$_2$, modeled with the screened-Coulomb
potential of Nakano {\it et al.}~\cite{vashishta90,nakano94} which has
been shown to give realistic structures and a good account of
a number of dynamical properties.  We prepare these runs
starting from randomly packed unit cells, and relax them through 5000
ART iterations. This procedure ensures absence of correlation, both
with the crystalline state as well as between runs. After relaxation, a
further 5000 ART iterations are performed on each cell. Slightly more than
half of these iterations show a clean convergence to a saddle point,
providing a database of 5645 events.  An analysis of these events can
give us a unique glimpse at the basic nature of activated mechanisms in
this material.

We checked for systematic effects caused by the initialization
procedure or by the potential used: a comparison with events from a
shorter run, starting from an MD-prepared 576-atom sample, indicates
that the nature of the events is independent of the preparation mode; a
comparison with events from a shorter run in which the van Beest
potential was used, with parameters as in \onlinecite{vollmayr},
indicates that, unless stated otherwise, the results presented here are
at least qualitatively similar between these potentials.

The efficiency of ART does not depend directly on the height of the
activation barrier or the complexity of the move.  The likelihood
for a particular event to be sampled by ART, however, is not clearly
related to the preferences of nature; entropic considerations, for
instance, are left out. This draw-back is not present in molecular
dynamics (MD).  However, the time scales accessible to MD are limited
by the phonon time scale: events in {\it g}-SiO$_2$ can only be
generated with some efficiency if the simulations are performed at
elevated temperatures of 4000 K or more \cite{litton97}; events sampled
at these high temperatures are likely to provide an incomplete
representation of those occurring below the glass transition.
Our approach is therefore to generate a whole distribution of events
with ART, and to obtain an overview of the possible types of activated
mechanisms in {\it g}-SiO$_2$, by classifying these in terms of energy,
defects or topological changes.  Further study of the details of the
energy landscape will be necessary in order to simulate the dynamics of
low temperature configurations.

During the events acquisition, the configurational energy decreases by
about 30 meV per atom. The density of coordination defects fluctuates
but does not show a clear trend. With the Nakano potential, the samples
have roughly the right density for {\it g}-SiO$_2$ and about one
percent of the bonds between O and Si are missing, compared to perfect
coordination.  The defects produced are almost uniquely dangling bonds;
we do not find any homopolar bond nor, consequently, any superoxide
radical or Frenkel pair~\cite{griscom86,hosono98}.  The van Beest
potential~\cite{beest90}, that we used for comparison, tends to produce
dense phases and its cut-off needs to be finely tuned to get the right
density.  Even in that case, it favors overcoordination~\cite{vollmayr}.

\begin{figure}
\epsfxsize=8cm
\epsfbox{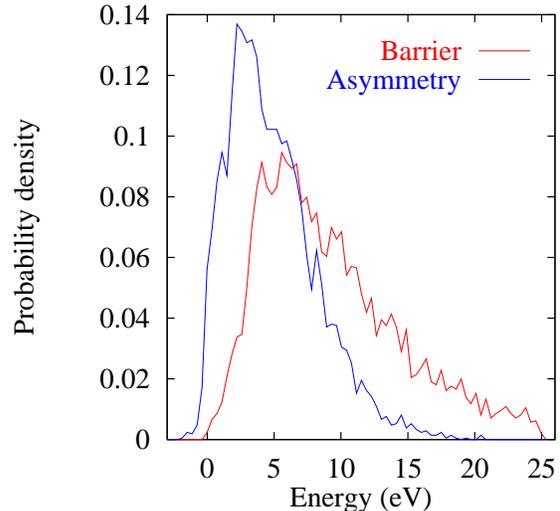}
\caption{Distributions of the energy barriers (red curve) and
asymmetries (blue curve), obtained from the database of all events.
Both distributions are continuous; no gap is seen. The barrier
distribution peaks around 5 eV and extends well beyond the
physically relevant values, but also displays a significant weight
below 5 eV. The asymmetry distribution is narrower, and peaks around 2.5
eV.  A small fraction of the events show a negative asymmetry, {\it
i.e.} lower the energy. One expects this fraction to be small, since
the configurations are well relaxed.  Because of the exponential nature
of the activation process, the presence of high energy barriers and
states is irrelevant because they are not sampled on normal time
scales; it is the low energy part of the spectrum which determines the
dynamics.
\label{fig:fullene}}
\end{figure}

We first look at properties averaged over the whole database. Figure
\ref{fig:fullene} shows broad and continuous distributions of the
activation barriers and the energy asymmetries (initial to final
minimum energy difference). We find that the height of the barrier and
the asymmetry of the well are only weakly correlated.

\begin{figure}
\epsfxsize=8cm
\epsfbox{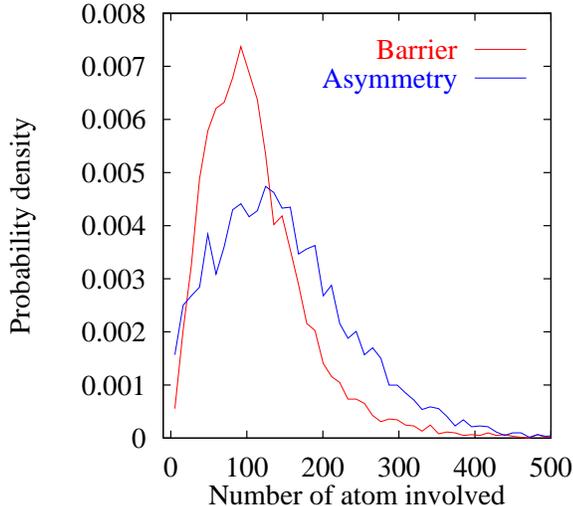}
\caption{Distribution of the number of atoms involved in events, {\it
i.e.}, the number of atoms that have moved more than
0.1~\AA, from the initial minimum to the saddle point
(red curve) and to the final minimum (blue curve).  The number of
displaced atoms increases as one moves from the saddle point to the new
minimum. An event involves typically between 50 and 250 atoms, but a
significant number of events involve up to 300 atoms and more. This
shows that relaxation in glasses is a collective phenomenon, in
agreement with experimental results by J{\'e}r{\^o}me and Commandeur
\protect\cite{jerome97}, but different from the related material {\it
a}-Si where only about 40 atoms are typically involved in events
\protect\cite{barkema98}.  The
size of these events puts a lower bound on the number of atoms
necessary in numerical simulations to avoid large finite-size effects.
\label{fig:fulln}}
\end{figure}

Besides the barrier, the spatial extend of events is another important
quantity.  For each event, we determine the number of atoms displaced
by more than a threshold distance $r_c$.  The number of atoms
participating in an event depends of course on the value
of this threshold. Typically, an event is accompanied by a local volume
contraction or expansion. In elastic media, the displacement of the
surrounding atoms decreases quadratically away from the center of the
event. The number of atoms moving more than a cut-off distance $r_c$
will therefore decrease as ${r_c}^{-3/2}$, as long as $r_c$ is in the
elastic regime, and the number of atoms much smaller than the sample
size; in our case, this scaling is obeyed between 0.05 and 1 \AA.
In figure \ref{fig:fulln} we plot this distribution for a threshold
of $r_c=0.1$ \AA, the typical vibrational amplitude of silicon
at room temperature. As can be seen from this figure, events typically
involve the motion of hundreds of atoms with simultaneous diffusion of
both species to varying degrees: diffusion therefore should not be
thought of in terms of elemental jumps but of complex rearrangements.

In terms of correlation, larger events are not found to require a
higher activation energy: size and energetics are almost entirely
uncorrelated.  Moreover, no correlation is found between the distance
by which Si or O move and the corresponding activation barriers; Si and
O have thus the same activation energy.

These results provide a consistent picture for macroscopic diffusion.
Theoretical work by Limoge~\cite{limoge90b} suggests that the
effective activation energy should be around the maximum of its
distribution, {\it i.e.}, here, about 5 eV~\cite{energy}. A similar
activation energy was found by Litton and Garofalini for O and Si in
MD simulations of molten silica, although the diffusion mechanism
described is different from what we see, probably due to the high
temperature of the simulations (from 4800 to 7200 K)\cite{litton97}.
Experiments report activation energies of 6.0 eV for
Si~\cite{brebec80}, obtained in electrically fused quartz, and 4.7 eV
for O, with a much smaller prefactor~\cite{mikkelsen84}, obtained in
vapor-phase deposited amorphous silica.  The difference in
experimental activation energies might be caused by the different
sample preparation techniques, resulting in different types of
impurities~\cite{litton97}.

More microscopic information on the nature of the events can be
obtained by studying the topology of the network. For this purpose, we
divide the events into three distinct categories: {\it perfect events}
where only perfectly coordinated atoms change neighbors,
{\it conserved events} that involve only diffusion of coordination defects
(dangling and floating bonds) and events that create or anneal
coordination defects.

Amorphous Si and {\it g}-SiO$_2$ are thought to be conceptually
similar, both described by Zachariasen's continuous random networks.
However, while perfect events play a central role for both relaxation
and self-diffusion in {\it a}-Si~\cite{barkema98}, they are rare in
{\it g}-SiO$_2$: the strong ionicity of SiO$_2$ enforces chemical
ordering, so that atomic exchanges have to occur at the second neighbor
level, inducing more strain or  larger topological rearrangements than
in {\it a}-Si. Perfect events comprise only about one percent of the
total number produced. A third of these events involves local
topological rearrangements, mostly two Si exchanging a pair of
neighboring O. Such moves can only happen with a relatively low energy
barrier if the local rigidity of the network is reduced by nearby
undercoordinated atoms.  Two thirds of the perfect events do not involve a
topological modification but simply some slight local rearrangement,
with displacements on the order of $0.1$ \AA \, and asymmetries of
about $10^{-4}$ eV. Such events could be candidates for tunneling
states. 

We find 906 conserved events, i.e., events describing the diffusion of
defects. Such defects are almost exclusively dangling bonds, on both Si
($E'$ centers) and O (non-bridging oxygens), although a few highly
energetic floating bonds on O are also present. We see no sign of point
defects, which would show themselves by a strong spatial correlation
between dangling or floating bonds. Events describe overwhelmingly
single-dangling-bond diffusion mechanisms. The simplest of these is a
jump of a dangling bond from one atom to its neighbor, an example of
which is given in Figure \ref{fig:danglingdiff}.  More complex events are
also seen, involving jumps to the second or third neighbor, or local
rearrangements along a loop. All these mechanisms have relatively well
defined barriers and asymmetries. From their statistics we can obtain
structural information. For instance, a comparison of near-neighbor
dangling bond diffusion involving different topological
rearrangements shows that the average cost of creating a 3-fold ring in
silica is $1.5 \pm 0.2$ eV. This value is larger than the $0.25-0.81$
eV of {\it ab-initio} calculations on fully relaxed
molecules~\cite{okeefe84,hamann97}, suggesting that the local
strain on the network caused by topological disorder can
affect significantly their effective energy.

More than 80 percent of the events produced involve the creation or the
annihilation of coordination defects, with a wide spectrum of energies
and configurations. Events with a low barrier and asymmetry, the ones
determining the dynamics, are often topologically simple, like the
annihilation of one or two pairs of dangling bonds or their creation.
In effect, the creation (or annihilation) of a pair of defect costs
(saves) much less energy than would be naively thought by simply
considering the breaking of a bond in a crystal or a molecule:  the
elastic energy stored in the network will often counter the bonding
energy.  Contrary to what is found in crystalline silica, the creation
of a defect in the glass can have an activation energy and asymmetry
that is comparable to those associated with their diffusion.  For
example, creating a pair of dangling bonds in order to remove a 3-fold
ring costs only about 0.4 eV, much less than what would be expected
in an unstrained environment.

The above results provide the following picture regarding relaxation
and diffusion in {\it g}-SiO$_2$.  Mechanisms responsible for
relaxation and diffusion in {\it g}-SiO$_2$ are the creation, diffusion
and annihilation of coordination defects, and can require the collective
displacement of hundreds of atoms.  The types of defects that
dominate the dynamics are dangling bonds, either attached to a Si atom
($E'$ centers), or to an O atom (non-bridging oxygens); a pair of these
defects can easily be created and annealed, with an activation energy
that is often similar to what is required for the diffusion of these
defects. Moreover, all these mechanisms involve O and Si with almost
equal weight, indicating that the two species should diffuse with
roughly the same activation barrier.  These elementary mechanisms are
fundamentally different from those found in amorphous silicon, which
underlines the rich diversity in the microscopic dynamics of network
glasses.

\noindent {\em Acknowledgements.}  Part of the calculations were
carried out on the CRAY T3E of HPAC.  This work is supported in part by
the ``Stichting voor Fundamenteel Onderzoek der Materie (FOM)'', which
is financially supported by the ``Nederlandse Organisatie voor
Wetenschappelijk Onderzoek (NWO)'', and by the NSF under grant
DMR 9805848.

\bibliographystyle{prsty}

\begin{thebibliography}{99}

\bibitem{griscom86} D.L. Griscom,
Mat. Res. Soc. Symp. Proc. {\bf 61}, 213-221 (1986).

\bibitem{limoge90} Y. Limoge, {\it in Diffusion in Materials},
A. L. Laskar {\it et al.} (eds.), Kluwer Academic Publishers, 601 (1990).

\bibitem{lamkin92}  M.A. Lamkin, F.L. Riley and R.J. Fordham,
J. Europ. Ceram. Soc. {\bf 10}, 347-367 (1992).

\bibitem{stillinger98} S. Sastry,  P.G. Debenedetti, and F.H. Stillinger,
{\em Nature} {\bf 393},  554-557 (1998).

\bibitem{art} G.T. Barkema and N. Mousseau,
Phys. Rev. Lett. {\bf 77}, 4358-4361 (1996).

\bibitem{artb} N. Mousseau and G.T. Barkema, Phys. Rev. E {\bf 57},
2419-2424 (1998).

\bibitem{vashishta90} P. Vashishta, R.K. Kalia, J.P. Rino and
I. Ebbsj{\"o}, Phys. Rev. B {\bf 41}, 12197-12209 (1990).

\bibitem{nakano94} A. Nakano, L. Bi, R.K. Kalia and P. Vashishta,
Phys. Rev. B {\bf 49}, 9441-9452 (1994).

\bibitem{beest90} B.W.H. van Beest, G.J. Kramer, and R.A. van Santen,
Phys. Rev. Lett. {\bf 64}, 1955-1958 (1990).

\bibitem{vollmayr} K. Vollmayr, W. Kob and K. Binder,
Phys. Rev. B {\bf 54}, 15808-15827 (1996).

\bibitem{litton97} D.A. Litton and S.H. Garofalini,
J. Non-Cryst. Sol. {\bf 217}, 250-263 (1997).

\bibitem{hosono98} H. Hosono, H. Kawazoe and N. Matsunami,
Phys. Rev. Lett. {\bf 80}, 317-320 (1998).

\bibitem{jerome97} B. J{\'e}r{\^o}me and J. Commandeur,
{\em Nature} {\bf 386}, 589-592.

\bibitem{limoge90b} Y. Limoge and J.L. Bocquet,
Phys. Rev. Lett. {\bf 65}, 60-63 (1990).

\bibitem{energy} The exact value of the average energy barrier depends
closely on the potential used; any comparison with other theoretical
and experimental results must therefore be made with some care. What
should be noted here is that (1) we find the same barrier for Si- and
O-dominated events and (2) the energies are compatible with other
simulations and experiments.

\bibitem{brebec80} G. Br{\'e}bec, R. S{\'e}guin, C. Sella, J. Bevenot,
and J.C. Martin, Acta Metall. {\bf 28}, 327- (1980).

\bibitem{mikkelsen84} J.C. Mikkelsen,
Appl. Phys. Lett. {\bf 45}, 1187-1189 (1984).

\bibitem{barkema98} G.T. Barkema and N. Mousseau,
Phys. Rev. Lett. {\bf 81}, 1865 (1998).

\bibitem{okeefe84} M. O'Keefe and G.V. Gibbs,
J. Chem. Phys. {\bf 81}, 876 (1984).

\bibitem{hamann97} D.R. Hamann,
Phys. Rev. B {\bf 55}, 14784-14793 (1997).

\end{thebibliography}

\end{document}